\documentclass[12pt]{iopart}

\usepackage{graphicx}
\begin{document}

\title[Anomalous localization enhancement in 1D non-Hermitian disordered lattices]{Anomalous localization enhancement in one-dimensional non-Hermitian disordered lattices}

\author{Ba Phi Nguyen$^1$, Duy Khuong Phung$^2$ and Kihong Kim$^{3}$\footnote{Author to whom any correspondence should be addressed}}

\address{$1$
Department of Basic Sciences, Mientrung University of Civil Engineering, Tuy Hoa 620000, Vietnam\\$2$ Computing Fundamentals Department, FPT University, Hanoi 100000, Vietnam\\$3$  Department of Energy Systems Research and Department of Physics, Ajou University, Suwon 16499, Korea}
\ead{khkim@ajou.ac.kr}
\vspace{10pt}

\begin{abstract}
We study numerically the localization properties of eigenstates in a one-dimensional random lattice described by a non-Hermitian disordered Hamiltonian, where both the disorder and the non-Hermiticity are inserted simultaneously in the on-site potential. We calculate the averaged participation number, the Shannon entropy and the structural entropy as a function of other parameters. We show that, in the presence of an imaginary random potential, all eigenstates are localized in the thermodynamic limit and strong anomalous Anderson localization occurs at the band center. In contrast to the usual localization anomalies where a weaker localization is observed, the localization of the eigenstates near the band center is strongly enhanced in the present non-Hermitian model. This phenomenon is associated with the occurrence of a large number of strongly-localized states with pure imaginary energy eigenvalues.
\end{abstract}

%

%
%
%
%
\section{Introduction}
\label{sec:1}
The problem of multiple scattering and localization of waves has been among the most difficult and fascinating topics in the study of wave propagation in disordered systems. An ubiquitous feature of wave propagation in such systems is the spatial confinement of waves called Anderson localization, which occurs due to the destructive interference in the forward direction. It was first predicted theoretically for electron scattering in disordered solids and later extended to classical waves \cite {And,Lee,She}. Even though it took a long time for the theory of Anderson localization to be accepted widely by researchers, by now this phenomenon has been observed experimentally in various wave systems including light waves \cite{Sch,Lah} and matter waves \cite{Bil,Roa}. A brief overview of some remarkable experimental results has been given in \cite{Aspect} as well.

The theory of Anderson localization in one-dimensional (1D) random systems has been developed extensively. One of the key characteristic quantities of the theory is the localization length that determines the length scale over which the envelope of the localized wave function exponentially decays. Based on a non-degenerate perturbation theory, an approximation analytical expression for the localization length was derived in the limit of weak and uncorrelated disorder \cite{Thouless1}. Shortly afterwards, however, the numerical calculations performed in \cite{Czycholl} have indicated that this expression fails for energies near the band center. For the purpose of resolving this discrepancy, which was termed the band-center anomaly, a degenerate perturbation theory around the band center was developed in \cite{Kappus}. Using this theory, an analytical expression for the localization length, which explained the anomalous localization behavior at the band center, was derived. By applying different approaches, a complete analytical solution regarding this problem has been found later \cite{Kravtsov1,Kravtsov2,Tessieri1}. More recently, the study of the band-center anomaly has been extended to quasi-1D Anderson models \cite{Nguyen1} and 1D Anderson models with correlated disorder \cite{Ramola,Tessieri2} as well. In addition to the band-center anomaly, similar anomalies have been found near the band edges \cite{Deych1,Deych2,Hern} as well as at other spectral points \cite{Derrida,Alloatti,Sepehrinia1,Sepehrinia2}. We emphasize that all these anomalies have been observed for Anderson models with real-valued on-site potentials. One of the main findings has been that the anomaly leads to an unusual enhancement of the localization length, namely, it weakens Anderson localization \cite{Izrailev,Krimer,Amir}.

Since it is generally believed that the onset of Anderson localization requires multiple scattering from the random fluctuations of the real-valued potential, most of the well-established results in this field have been obtained for Hermitian systems. In recent years, however, there has been considerable attention to wave propagation and localization in non-Hermitian systems both without and with disorder \cite{Fre,Paa,Asa,Jia,Hat,Sil,Kal,Bas,Vaz,Mej,Kar,Nguyen2}. Unlike in the case of Hermitian systems where disorder-induced localization has been understood quite well, there exist debates about the localization behaviors in non-Hermitian models. Remarkably, it has been shown that the presence of either amplification or absorption in the system under consideration gives rise to the same degree of suppression of the transmittance \cite{Fre,Paa,Asa}. However, Jiang {\it et al} have argued that this conclusion is an artifact originating from time-independent calculations \cite{Jia}. Using the time-dependent Maxwell equation, they have demonstrated that above a certain critical length scale, the amplitudes of both reflected and transmitted waves diverge in the presence of amplification. In order to overcome this divergence problem, Basiri {\it et al} \cite{Bas} have used a different setup which was proposed in \cite{Christo} to revisit this problem. From the numerical and approximate analytical calculations, these authors have concluded that the presence of disorder only in the imaginary part of the potential can lead to localization.
They have argued that the localization arises due to a balance between gain (or loss) and diffraction and its mechanism is qualitatively
different from the Anderson model with disorder only in the
real part of the potential.
Besides, they have also pointed out that there exists some similarity in the scaling behavior between the system under consideration and the standard Hermitian Anderson model. A natural question we ask is whether the anomalous localization behavior at the band center occurs in the non-Hermitian case with an imaginary random potential or not. This problem, to the best of our knowledge, has not been considered in the literature. This provides us with the motivation for pursuing this research direction.

\section{Theoretical Model}
\label{sec:2}
\subsection{Model}

\begin{figure}
\centering
\includegraphics[width=12cm]{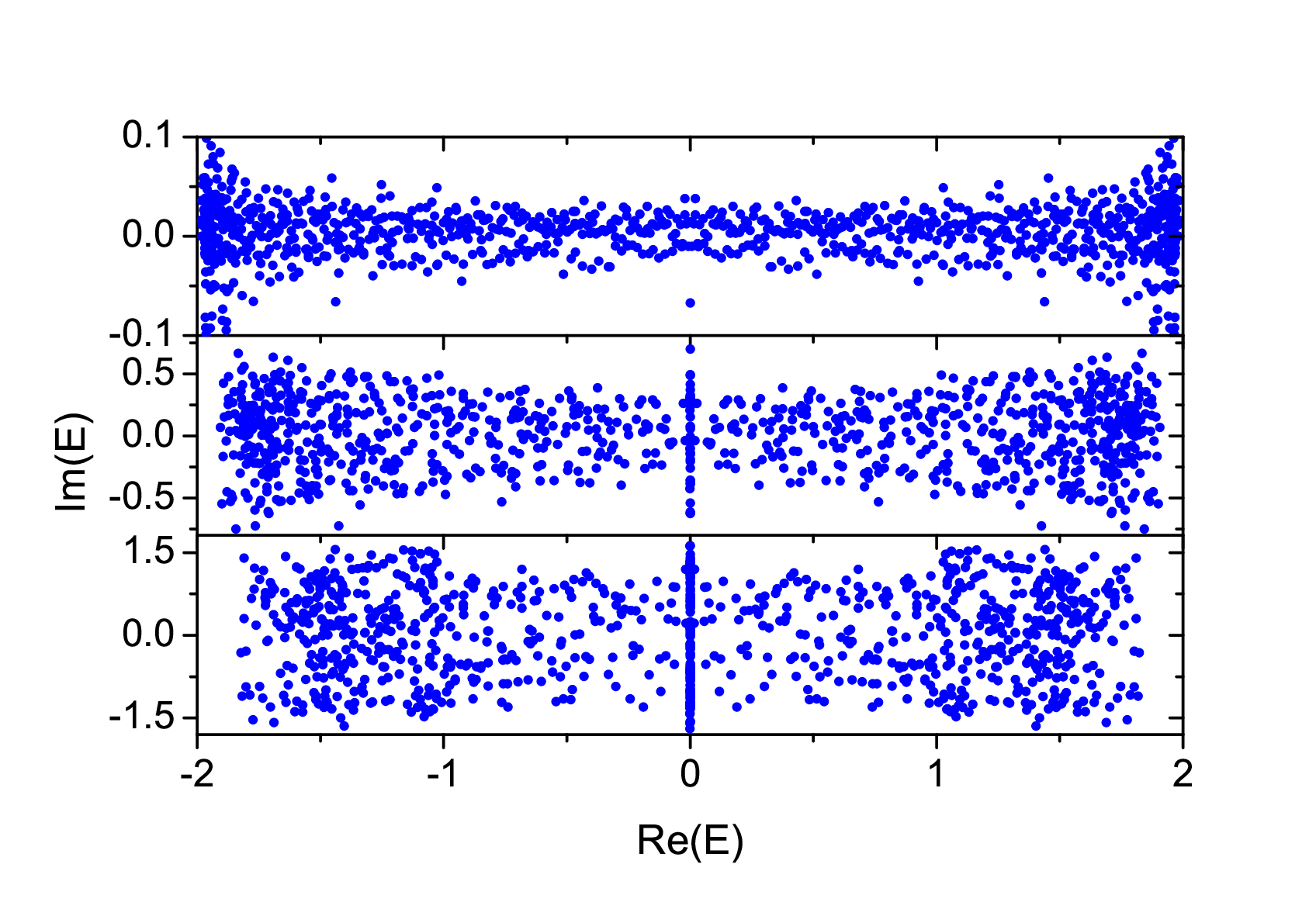}
\caption{Real and imaginary parts of the energy eigenvalues for the system under consideration, when $\epsilon_{n}^{\rm R}$ is zero and $\epsilon_{n}^{\rm I}$ is selected randomly from a uniform distribution in the interval $[-W/2, W/2]$ with  $W=0.5$, 2 and 4 (from the top panel to the bottom panel). The system size is $N=1000$.}
\end{figure}

Let us consider an array of $N$ coupled optical waveguides with a constant tunneling amplitude $V$ with which light is transferred from waveguide to waveguide. In the approximation where only the coupling between nearest-neighbor waveguides is considered, the wave propagation process in such a system can be modeled by a set of coupled discrete Schr\"{o}dinger equations for the field amplitudes $C_{n}$ given by
\begin{eqnarray}
i \frac{dC_{n}}{dz}=-V\left(C_{n-1}+C_{n+1}\right) + \epsilon_{n}C_{n},
\label{equation1}
\end{eqnarray}
where $z$ is the paraxial propagation distance, $n$ ($=1,2,\cdots, N$) is the waveguide number, and $\epsilon_{n}$ is the on-site potential \cite{Christo}.
The stationary solutions of equation~(\ref{equation1}) can be represented in the conventional form, $C_{n}(z)=\psi_{n}e^{-iEz}$, with $E$ as the energy of an eigenstate $n$. Then we obtain the stationary discrete Schr\"{o}dinger equation of the form
\begin{eqnarray}
 E\psi_{n}=-V\left(\psi_{n-1}+\psi_{n+1}\right)+\epsilon_{n}\psi_{n}.
\label{equation2}
\end{eqnarray}

In the present work, the on-site potential $\epsilon_{n}$ is considered to be complex. Specifically, it takes the form,
\begin{eqnarray}
\epsilon_{n}=\epsilon_{n}^{\rm R}+i\epsilon_{n}^{\rm I},
\label{equation3}
\end{eqnarray}
where the real part $\epsilon_{n}^{\rm R}$ is chosen to be the same for all lattice sites (below we put $\epsilon_{n}^{\rm R}=0$ for simplicity), whereas the imaginary part $\epsilon_{n}^{\rm I}$ is assumed to be random, uncorrelated and distributed uniformly  in the finite interval $[-W/2,W/2]$, where $W$ measures the strength of disoder. When the random on-site potential is complex, the energy eigenvalues are no longer real but become complex. An example of the distribution of the eigenvalues in the complex plane is shown in figure~1. It has some interesting features that are worth mentioning. When $W$ is sufficiently small ($W=0.5$, for example),
the imaginary parts of the eigenvalues show larger fluctuations near the band edges.
In this case, an approximate analytical expression describing the envelope of the points was derived in \cite{Izyumov1} and \cite{Izyumov2}.
As we increase $W$, the number of eigenstates close to the band edges decreases gradually. At the same time, the number of eigenstates with pure imaginary energy eigenvalues (namely, ${\rm Re}(E)=0$) increases rapidly.
As we will show later, this feature plays a crucial role in explaining the localization behavior in the system under consideration.

\subsection{Participation number and information entropy}

The localization properties of disordered systems can be characterized in terms of various physical quantities such as the participation number, the Shannon entropy, the localization length and the intensity-intensity correlation function. In this study, we will focus on the participation number and the Shannon entropy, which can be defined for both finte and infinte systems. Specificially, for the $k$-th eigenstate $(\psi_{1}^{(k)},\psi_{2}^{(k)},\cdots,\psi_{N}^{(k)})^{\rm T}$ with the corresponding eigenvalue $E_{k}$, the participation number $P(E_{k})$ is defined by
\begin{eqnarray}
P(E_{k})=\frac{\left(\sum_{n=1}^{N}\big\vert\psi_{n}^{(k)}\big\vert^2\right)^2}{\sum_{n=1}^{N}\big\vert\psi_{n}^{(k)}\big\vert^4},
\label{equation4}
\end{eqnarray}
which approximately gives the number of lattice sites to which the $k$-th eigenstate extends \cite{Thouless}.

The participation number estimates the degree of spatial extension or localization of eigenstates. For a finite-size system, $P$ increases with increasing the system size $N$ in the extended regime. The completely extended state which spreads over the entire system uniformly corresponds to $P=N$. On the other hand, localized states exhibit much smaller values (in comparison with $N$) of $P$, which converge to constant values as $N\to \infty$. The most strongly localized state corresponds to $P=1$. Therefore, the participation number is bounded within the range $1\le P \le N$.
The inverse of $P$ called the inverse participation number is also frequently used \cite{Wegner}. It has been shown that the participation number is closely related to the localization length, $\xi$, which measures the exponential decay rate of a localized eigenstate in the thermodynamic limit. In the vicinity of the band center, the value of $P$ is almost identical to that of $\xi$ in the weak disorder limit  \cite{Krimer}. In general, the participation number represents an upper bound for the localization length ($P\geq\xi$).

In addition to $P$, we will also calculate the Shannon entropy, which is another basic quantity for the description of localized single-mode states in disordered systems. The Shannon entropy is defined by \cite{Izrailev2, Oliveira, Bogomolny}
\begin{eqnarray}
S(E_{k})=-\sum_{n=1}^{N}\vert\phi_{n}^{(k)}\vert^2\ln\left(\vert\phi_{n}^{(k)}\vert^2\right),
\label{equation5}
\end{eqnarray}
where $\phi_{n}^{(k)}$ represents the normalized amplitude of the $k$-th eigenstate wave function at site $n$,  $\phi_{n}^{(k)}=\psi_{n}^{(k)}/(\sum_{n=1}^{N}\vert\psi_{n}^{(k)}\vert^2)$. This quantity is essentially the logarithm of the number of lattice sites where the $k$-th eigenstate populates significantly. If all of the lattice sites are equally populated by a certain eigenstate, this state is fully delocalized and the corresponding Shannon entropy is $S=\ln N$. On the other hand, the most strongly localized state is the one resided significantly at only one lattice site and $S$ is equal to zero. Therefore the Shannon entropy is bounded as $0\le S\le \ln N$. Like the participation number, the Shannon entropy is closely related to the localization length. If an eigenstate is exponentially localized around a certain lattice site with the localization length $\xi$ $(1\ll \xi \ll N)$, then the relationship between $S$ and $\xi$ obeys the expression $S=1+\ln\xi+\rm{O}(1/\xi)$ \cite{Izrailev2}. This implies that the quantity $\exp(S)$ can be identified approximately as the localization length $\xi$.

The finite-size scaling analysis of $P$ as well as of $S$ gives a very useful information about the localized or extended nature of the eigenstates. In order to obtain the eigenvalues $E$  and eigenstates $\psi=(\psi_{1},\psi_{2},\cdots,\psi_{N})^{\rm T}$, which are used to calculate $P$ and $S$,  we numerically solve the eigenvalue problem
\begin{eqnarray}
\hat{H}\psi=E\psi,
\label{equation6}
\end{eqnarray}
where $\hat{H}$ is the random Hamiltonian matrix
\begin{eqnarray}
\hat{H} = \left( \begin{array}{cccccc}
			\epsilon_{1}		&-1				&0		&\cdots 		&0		&0\\
			-1				&\epsilon_{2}	&-1		&\cdots 		&0		&0\\
           0      &   -1    &      \epsilon_3    & \cdots     &    0   & 0\\
			\vdots 			&\vdots 			&\vdots 	&\cdots 		&\vdots 	&\vdots\\
			0				&0				&0		&\cdots 		&-1		&\epsilon_{N}
\end{array} \right)
\end{eqnarray}
constructed from equation~(\ref{equation2}) with the usual fixed boundary conditions (or hard-wall boundary conditions), $\psi_{0}=\psi_{N+1}=0$. Here, $V$ is taken to be unity with no loss of generality.

It is worth mentioning that the Schr\"odinger equation, equation~(\ref{equation1}), which describes the propagation of electromagnetic waves in the optical systems, is effectively the same as the equation which describes the transport of noninteracting electrons in the electronic systems. The key difference is that the evolution coordinate in the optical systems is the paraxial propagation distance, while it is replaced by the time in the electronic systems. Therefore, all the results obtained in this work are valid for electronic as well as optical systems.

\section{Numerical results}
\label{sec:3}

In this work, all quantities with the dimension of energy are measured in the unit of $V$, which is taken to be unity as mentioned before. The participation number $P$ and the Shannon entropy $S$ were obtained by averaging over the eigenstates with eigenvalues in a narow interval around a fixed ${\rm Re}(E)$ and ensemble averaging over 10000 different random configurations.

\begin{figure}
\centering
\includegraphics[width=12cm]{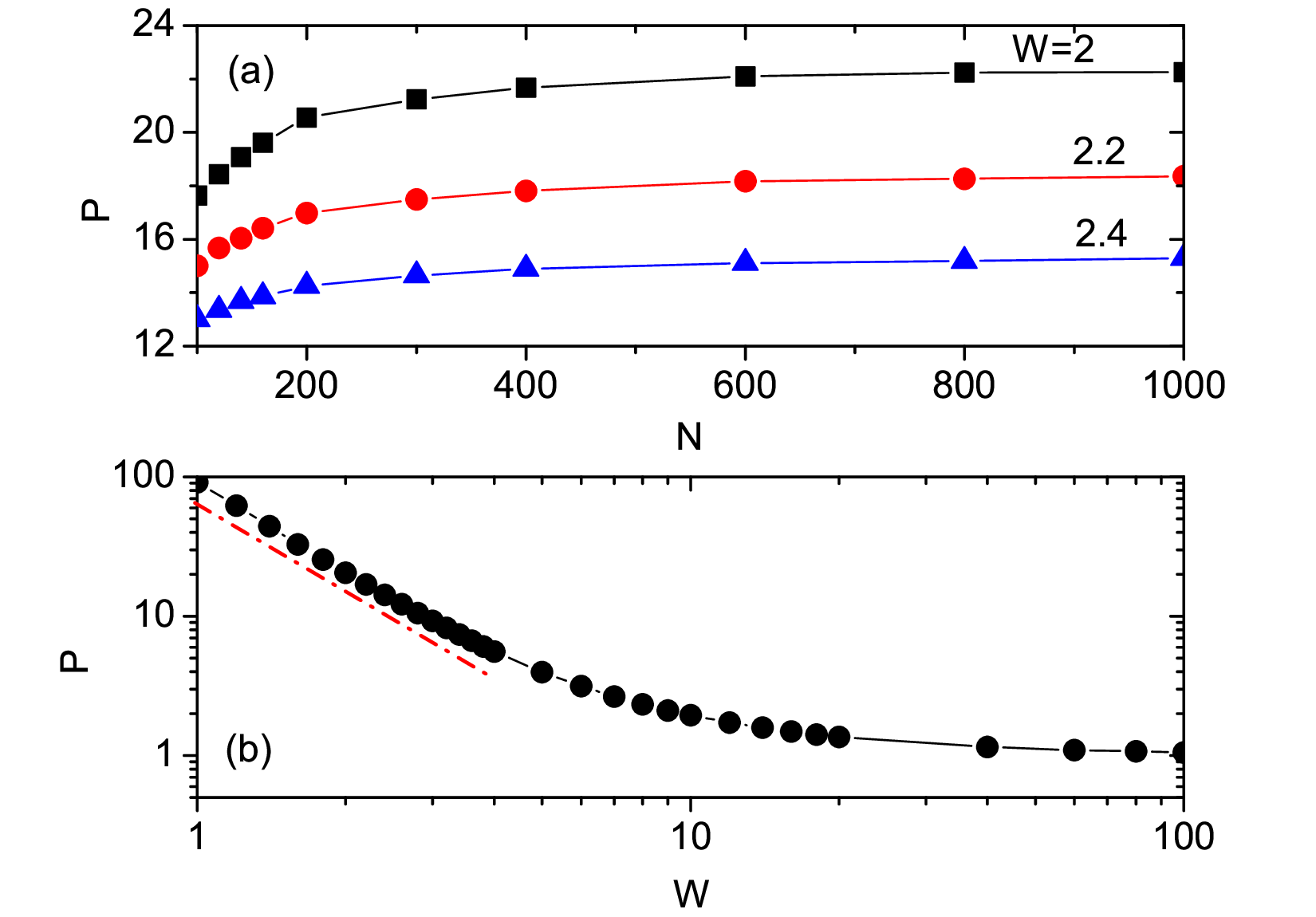}
\caption{(a) Participation number $P$ versus system size $N$ for different disorder strengths $W=2$, 2.2 and 2.4. (b) $P$ versus disorder strength $W$, when the system size is fixed to $N=1000$. The dash-dotted line indicates a power-law fit, $P(W)\propto W^{-2}$, to the data in the region $W< 4$. $P$ is obtained by averaging over the eigenstates with eigenvalues in the interval $[-0.1,0.1]$ around ${\rm Re}(E)=0$ and ensemble averaging over 10000 distinct disorder realizations.}
\end{figure}

In figure~2(a), we plot the participation number $P$, which was obtained by averaging over a small window around the band center, ${\rm Re}(E)\in [-0.1,0.1]$, versus $N$ for several typical values of the disorder strength $W$. For a given $W$, we find that $P$ initially increases as $N$ increases and rapidly approaches a saturation value $P_{s}$ at $N>N_{c}$. The value of $P_{s}$ is much smaller than the corresponding $N_{c}$. This indicates that the eigenstates close to the band center are localized states. We will see later that all states in the whole energy band are localized as well. Similarly to the standard Anderson model where the real part of the on-site potential is random, in the present model, the participation number is reduced, hence the localization is enhanced, as the disorder strength increases, as can be seen in figure~2(b). In the weak disorder limit, we find that the participation number is proportional to the disorder strength as a power law, $P\propto W^{-\alpha}$, with $\alpha=2$ (see the dash-dotted line in figure~2(b)). In the strong disorder limit, the participation number tends to $P=1$, indicating the occurrence of complete localization.

\begin{figure}
\centering
\includegraphics[width=12cm]{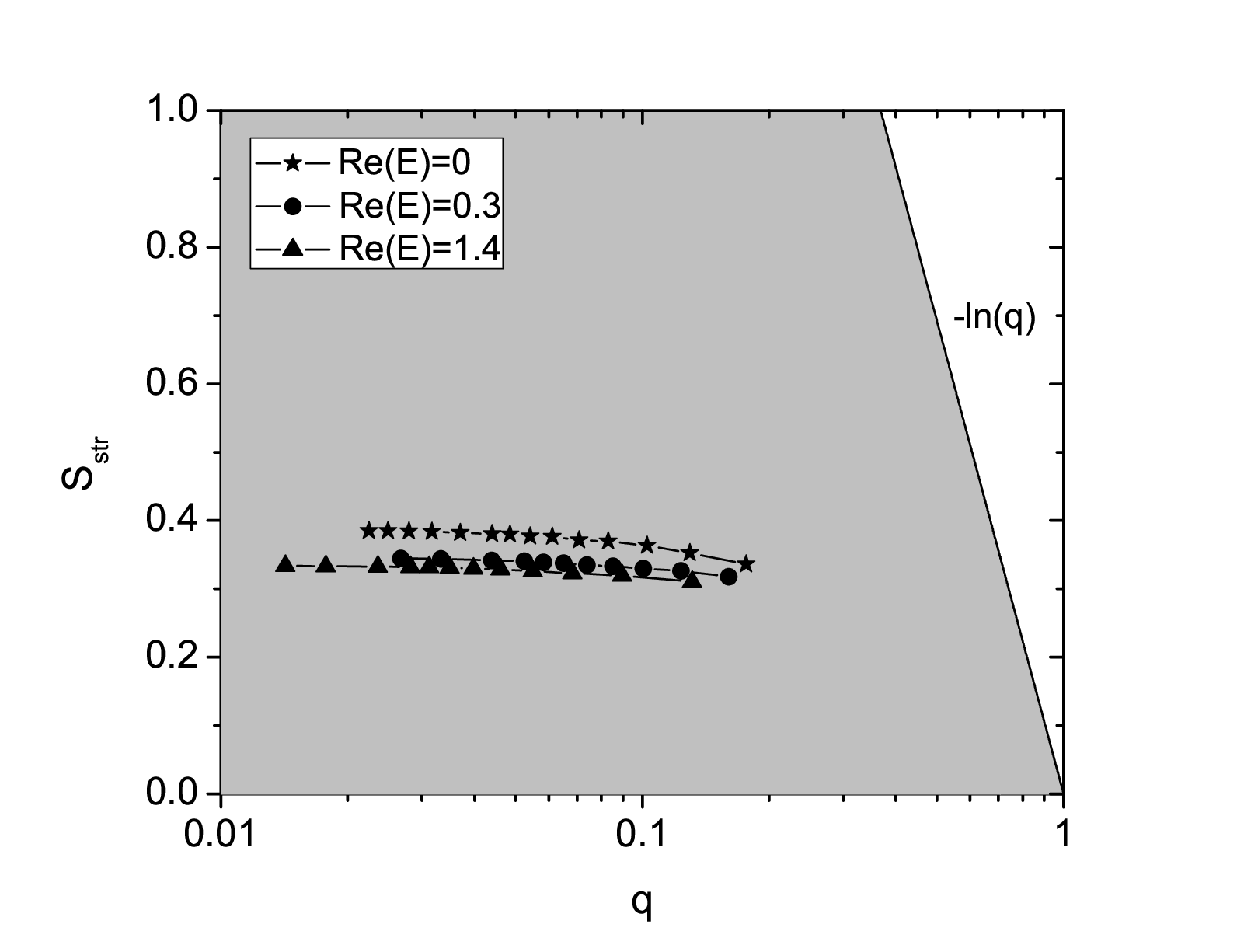}
\caption{Structural entropy $S_{\rm str}$ plotted versus spatial filling factor $q$ of eigenstates. The shaded area represents the permitted domain of the localization diagram, which is bounded by $0\le S_{\rm str}\le -\ln (q)$ with $0<q\le 1$. The symbols represent our numerical results for the eigenstates with ${\rm Re}(E)=0$, 0.3 and 1.4. Calculations were performed for the system size $N=1000$. The disorder strength is fixed to $W=2$.}
\end{figure}

The participation number is commonly used as a basic quantity for measuring the degree of localization. However, many localized states with completely different internal structures such as Gaussian, exponential and power-law decay forms can give rise to the same values of $P$. Therefore, we need a method to determine the shape of the localized eigenstate wave function explicitly. To this end, we use the quantity called structural entropy, which represents the shape of the localized wave function and is defined by \cite{Casati2,Pipek}
\begin{eqnarray}
S_{\rm str}(E_{k})=S(E_{k})-\ln P(E_{k}).
\label{equation7}
\end{eqnarray}
In \cite{Pipek}, it has been shown that there is a basic inequality between $S_{\rm str}$ and the spatial filling factor $q$ $(=P/N)$ given by $0\le S_{\rm str}\le -\ln (q)$ with $0<q\le 1$. This bound defines a permitted domain of the localization diagram, which has been proved to be universal. Therefore, the pair of the parameters $(q,S_{\rm str})$ should lie in this permitted domain for any generalized localized state. In figure~3, we plot $S_{\rm str}$ as a function of $q$ for several representative eigenstates with ${\rm Re}(E)=0$, 0.3 and 1.4. We see clearly that all points indicated by the pairs of the localization quantities $(q,S_{\rm str})$ are well included in the permitted region of the diagram. More specifically, we observe that apart from in the vicinity of ${\rm Re}(E)=0$ where $S_{\rm str}\sim 0.39$, we get $S_{\rm str}\sim 0.34$ for all other cases, as $q\to 0$ (or $N\to \infty$). These results are consistent with the theoretically predicted value for exponentially localized eigenstates, $S_{\rm str}=1-\ln 2\sim 0.31$, which was determined based on the continuous lattice model approach \cite{Pipek}. We believe that a substantial difference between the numerical result and theoretical prediction in the vicinity of the band center can be attributed to the existence of anomalous localization of eigenstates in that region.

\begin{figure}
\centering
\includegraphics[width=12cm]{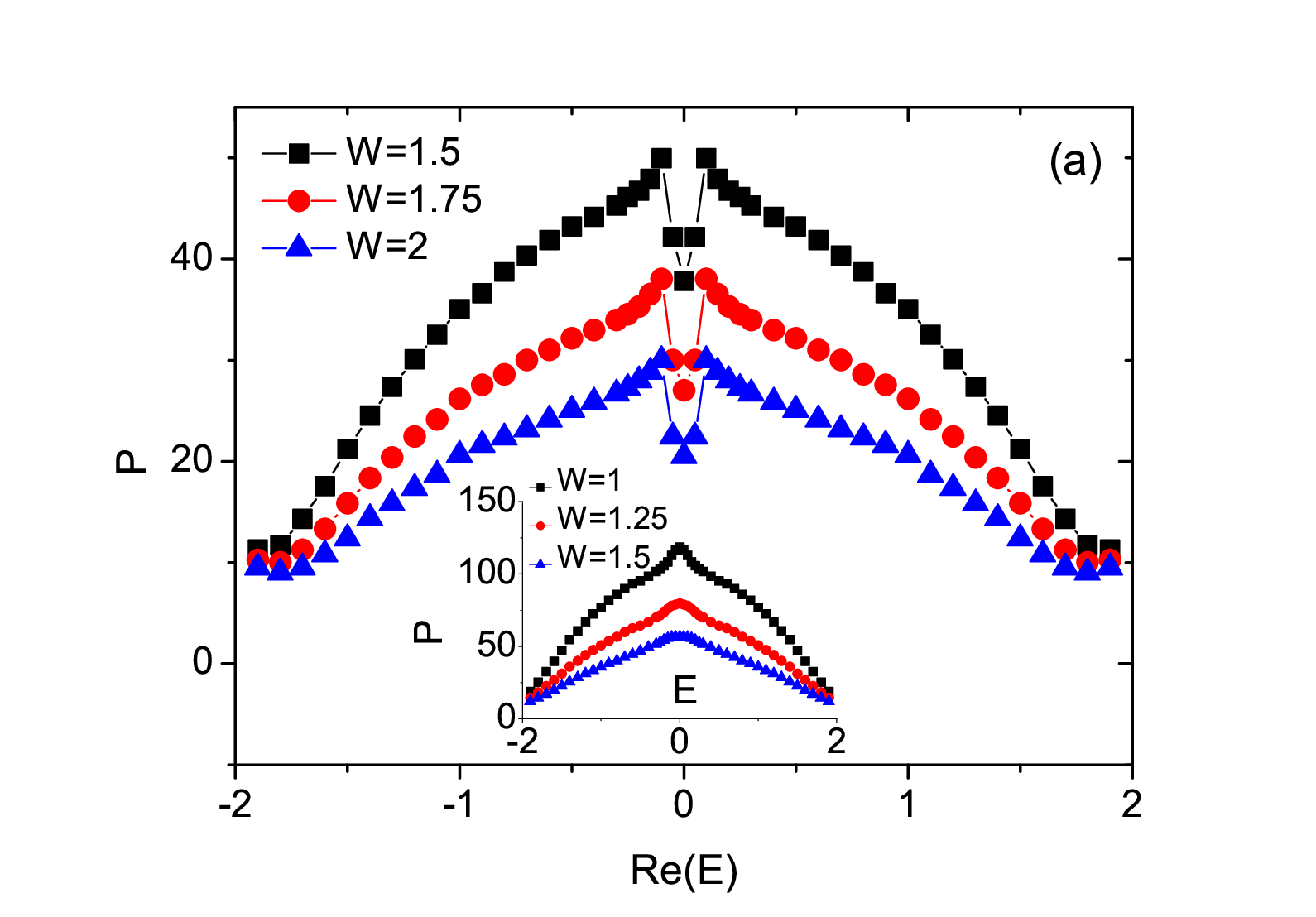}
\includegraphics[width=12cm]{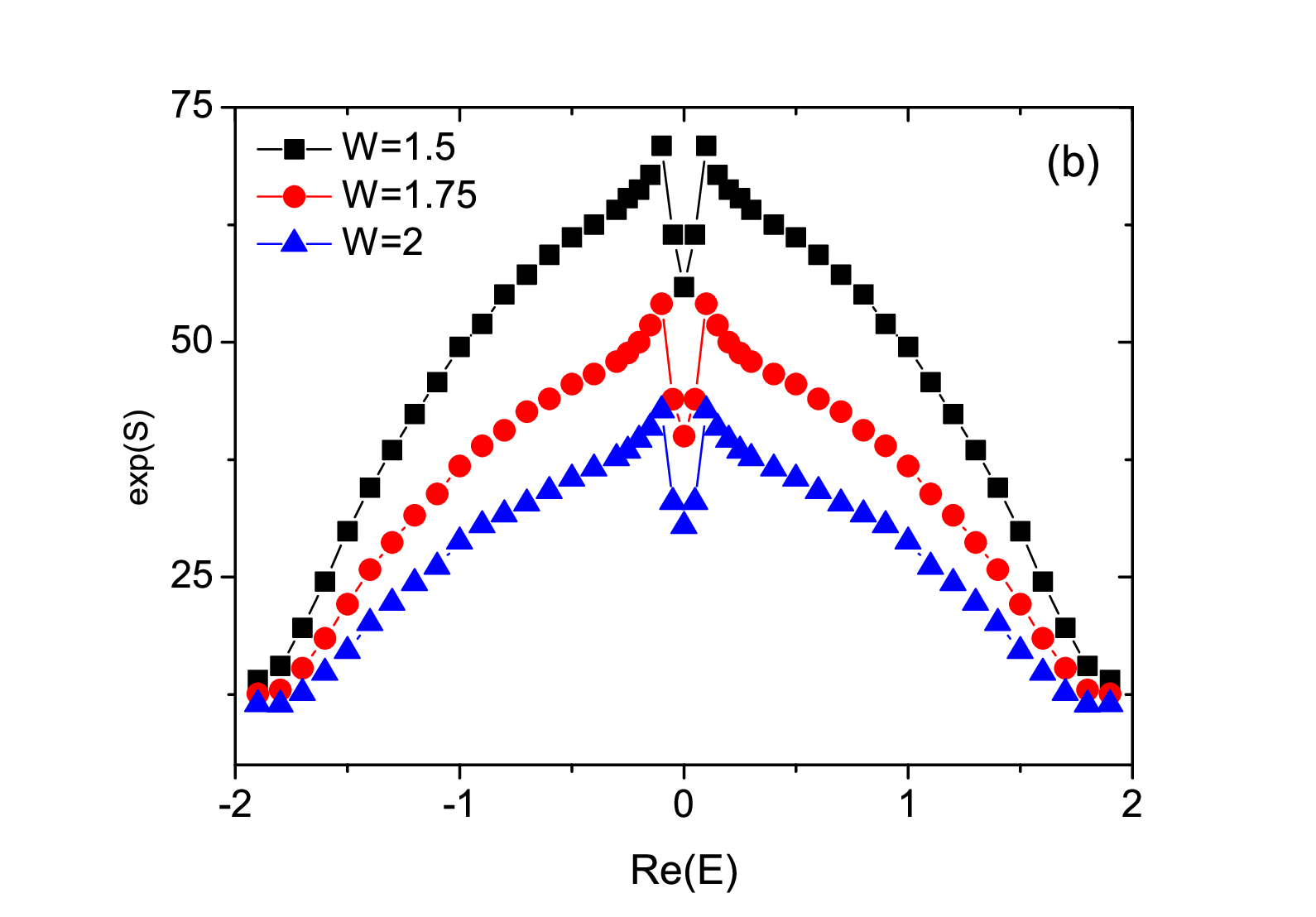}
\caption{(a) Participation number $P$ and (b) Shannon entropy $S$ plotted versus real part of the energy eigenvalue ${\rm Re}(E)$ for disorder strengths $W=1.5$, 1.75 and 2. $\exp(S)$ instead of $S$ is plotted for an easier comparison with $P$. For a given value of ${\rm Re}(E)$, $P$ and $S$ are obtained by averaging over the states in a small energy interval $\left[{\rm Re}(E)-0.05, {\rm Re}(E)+0.05\right]$. In the inset of figure~4(a), we also show the results obtained for the standard Anderson model with the corresponding real random potential when $W=1$, 1.25 and 1.5.}
\end{figure}

In figure~4, we show the participation number $P$ and the Shannon entropy $S$ as a function of the real part of the energy eigenvalue, ${\rm Re}(E)$, for disorder strengths $W=1.5$, 1.75 and 2, when the system size is fixed to $N=1000$. For the considered values of $W$, we see that the eigenstates close to the band edges are more strongly localized than those close to the band center as in the usual Anderson model. As we move away from the band edge towards the band center, $P$ and $S$ increase gradually until they decrease abruptly and exhibit {\it deep concave} regions around ${\rm Re}(E)=0$. This non-analytical dependence of $P$ and $S$ on ${\rm Re}(E)$ indicates an anomalous localization behavior of eigenstates. Remarkably, in contrast to the standard Anderson model in which one observes a sharp increase of the participation number (as illustrated in the inset of figure~4(a)), the present anomaly gives rise to a strong decrease of the participation number for eigenstates around ${\rm Re}(E)=0$. Moreover, in the former case, the band-center anomaly exists only when the disorder strength gets sufficiently small ($W\le 1$) \cite{Krimer}, whereas, in the present non-Hermitian case, a change in the disorder strength does not alter the height of the deep concave region, but brings about a change in the number of the eigenstates with ${\rm Re}(E)=0$ (see figure~5 below). We have also verified that the ratio $\exp(S)/P$ is always larger than unity for all values of ${\rm Re}(E)$. This is in good agreement with equation~(\ref{equation7}).

\begin{figure}
	\centering
	\includegraphics[width=12cm]{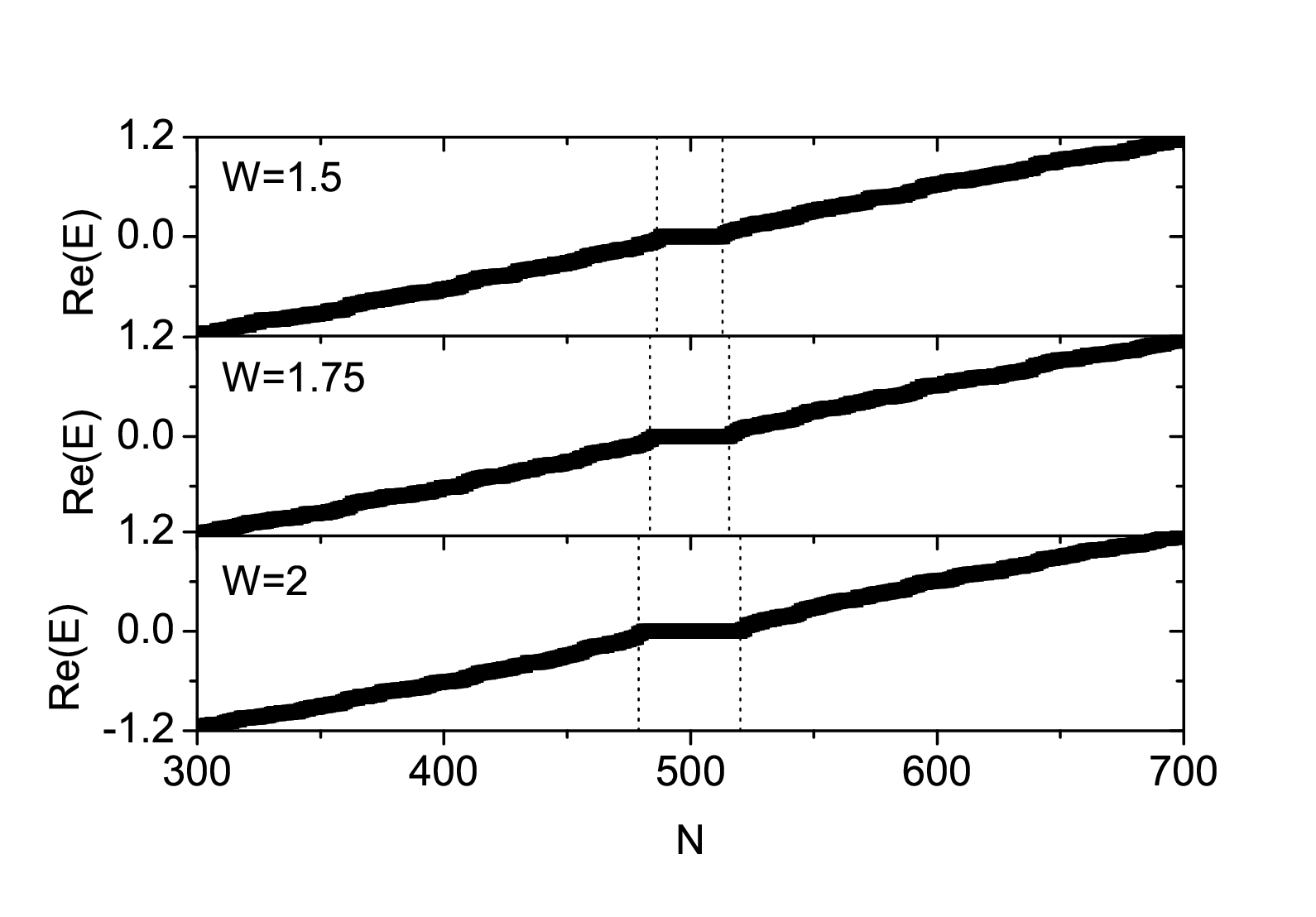}
	\caption{Real part of the energy eigenvalue versus eigenstate label arranged in order of increasing value of ${\rm Re}(E)$,
when the disorder strength $W$ is 1.5, 1.75 and 2. All the cases show that there exist a large number of eigenstates with pure imaginary eigenvalues. The number of such eigenstates increases with $W$.}
\end{figure}

In order to provide a quantitative explanation of the effect of anomalous localization enhancement which appears in 1D random non-Hermitian lattices, in figure~5, we show the real part of the energy eigenvalue versus eigenstate label arranged in order of increasing value of ${\rm Re}(E)$, for the identical parameters as in figure~4. From the numerical results, we find that the anomalous localization behavior at the band center is closely associated with the appearance of a large number of eigenstates with pure imaginary eigenvalues. If we consider only the real part of the eigenvalue to distinguish different eigenvalues, this is equivalent to having a large number of {\it degenerate} states with ${\rm Re}(E)=0$. From numerical calculations and approximate analytical treatments given below, we have verified that these states are strongly localized. Therefore, any superposition of these states results in strongly localized states. We have found that the number of such degenerate states increases with increasing $W$. As a consequence of this, in the strong disorder limit, the distribution function of ${\rm Re}(E)$ will exhibit a huge $\delta$ peak at ${\rm Re}(E)=0$.

In the strong disorder limit, the eigenstate wave function to the leading order in $V/W$ ($V/W\ll 1$) can be written as
\begin{eqnarray}
\psi_n=\delta_{nm}+VA\delta_{n,m-1}+VB\delta_{n,m+1},
\label{equation8}
\end{eqnarray}
where $A$ and $B$ are constants to be determined.
By substituting $n=m$, $m-1$, and $m+1$ respectively into equation (\ref{equation2}) with $\epsilon_{n}=i\epsilon_{n}^{\rm I}$, we obtain
\numparts
\begin{eqnarray}
E\psi_{m}&=&-V\left(\psi_{m-1}+\psi_{m+1}\right)+i\epsilon_{m}^{\rm I}\psi_{m},
\label{equation9a}\\
E\psi_{m-1}&=&-V\left(\psi_{m-2}+\psi_{m}\right)+i\epsilon_{m-1}^{\rm I}\psi_{m-1},
\label{equation9b}\\
E\psi_{m+1}&=&-V\left(\psi_{m}+\psi_{m+2}\right)+i\epsilon_{m+1}^{\rm I}\psi_{m+1}.
\label{equation9c}
\end{eqnarray}
\endnumparts
From equation (\ref{equation8}), we have
\begin{eqnarray}
\psi_m=1,~~
\psi_{m-1}=VA,~~
\psi_{m+1}=VB.
\label{equation10}
\end{eqnarray}
Substituting equation (\ref{equation10}) into equations (\ref{equation9b}) and (\ref{equation9c}), we obtain
\begin{eqnarray}
A=\frac{i}{\epsilon_{m}^{\rm I}-\epsilon_{m-1}^{\rm I}},~~B=\frac{i}{\epsilon_{m}^{\rm I}-\epsilon_{m+1}^{\rm I}}.
\label{equation11}
\end{eqnarray}
Finally, the energy eigenvalue to the second-order in $V$ is obtained by substituting these results into equation (\ref{equation9a}):
\begin{eqnarray}
E=i\epsilon_{m}^{\rm I}-i\left(\frac{1}{\epsilon_{m}^{\rm I}-\epsilon_{m-1}^{\rm I}}+\frac{1}{\epsilon_{m}^{\rm I}-\epsilon_{m+1}^{\rm I}}\right)V^2,
\label{equation12}
\end{eqnarray}
where $m$ is the position at which the wave function is localized and $V$ is set equal to unity in our study. In table~1, we show the numerical results for the wave function amplitude and the corresponding eigenvalue for several representative eigenstates with ${\rm Re}(E)=0$ obtained by solving equation~(\ref{equation6}) when $N=50$ and $W=10$ directly and compare the results with those obtained from the approximate analytical result, equation~(\ref{equation12}).
We find that the two results agree very well in this regime of strong disorder. We note that the eigenstates with pure imaginary eigenvalues seem to be favored to localize at lattice sites, where the corresponding values of $\epsilon_{m}^{\rm I}$ are large ($\sim \pm W/2$).

\begin{table}
\caption{\label{tab:table1} Numerical values of the wave function amplitude $\psi_{m}$ and the corresponding eigenvalue $E$ for several typical eigenstates in the strong-disorder region are listed. All eigenstates have ${\rm Re}(E)=0$. The used parameters are $N=50$ and $W=10$. The approximate values of ${\rm Im}(E)$ are obtained by equation~(\ref{equation12}).}
\footnotesize\rm
\begin{tabular*}{\textwidth}{@{}l*{15}{@{\extracolsep{0pt plus12pt}}l}}
\br
& state  &$m$ (localized center)&$\epsilon_{m}^{\rm I}$&$\vert\psi_{m}\vert$&${\rm Im}(E_{\rm exact})$&${\rm Im}(E_{\rm approx}$)\\
\mr
			&1&7&4.87183992&0.93916916&4.56421869&4.50660618 \\
			&2&1&4.66220070&0.98818589&4.52704191&4.52045924 \\
			&3&50&-4.57092533&0.99348626&-4.46465683&-4.46121117\\
			&4&29&-4.67401248&0.98511495&-4.43615195&-4.43805708\\
			&5&11&-4.84019334&0.94008911&-4.35074321&-4.39101455 \\
			&6&6&-4.55733493&0.98565603&-4.32524531&-4.32636897 \\
			&7&49&4.54366806&0.98127570&4.28462152&4.28389259\\
			&8&44&-4.48680400&0.94718793&-4.01643280&-4.04678787\\
\br
\end{tabular*}
\end{table}

From figure~4, we can easily notice that the localization of the eigenstates close to the band center is more strongly enhanced than that of those close to the band edges when the disorder strength becomes large. From this observation, we predict that, as the disorder strength gets sufficiently large, the eigenstates with ${\rm Re}(E)=0$ will become more strongly localized than all other eigenstates. This prediction is supported by the numerical results presented in figure~6, where several typical wave functions for the eigenstates with zero and non-zero real parts of the energy are shown. It is clearly seen that the eigenstates with pure imaginary eigenvalues (5th, 6th and 7th eigenstates in table 1) are more strongly localized than the remaining ones.

\begin{figure}
	\centering
	\includegraphics[width=12cm]{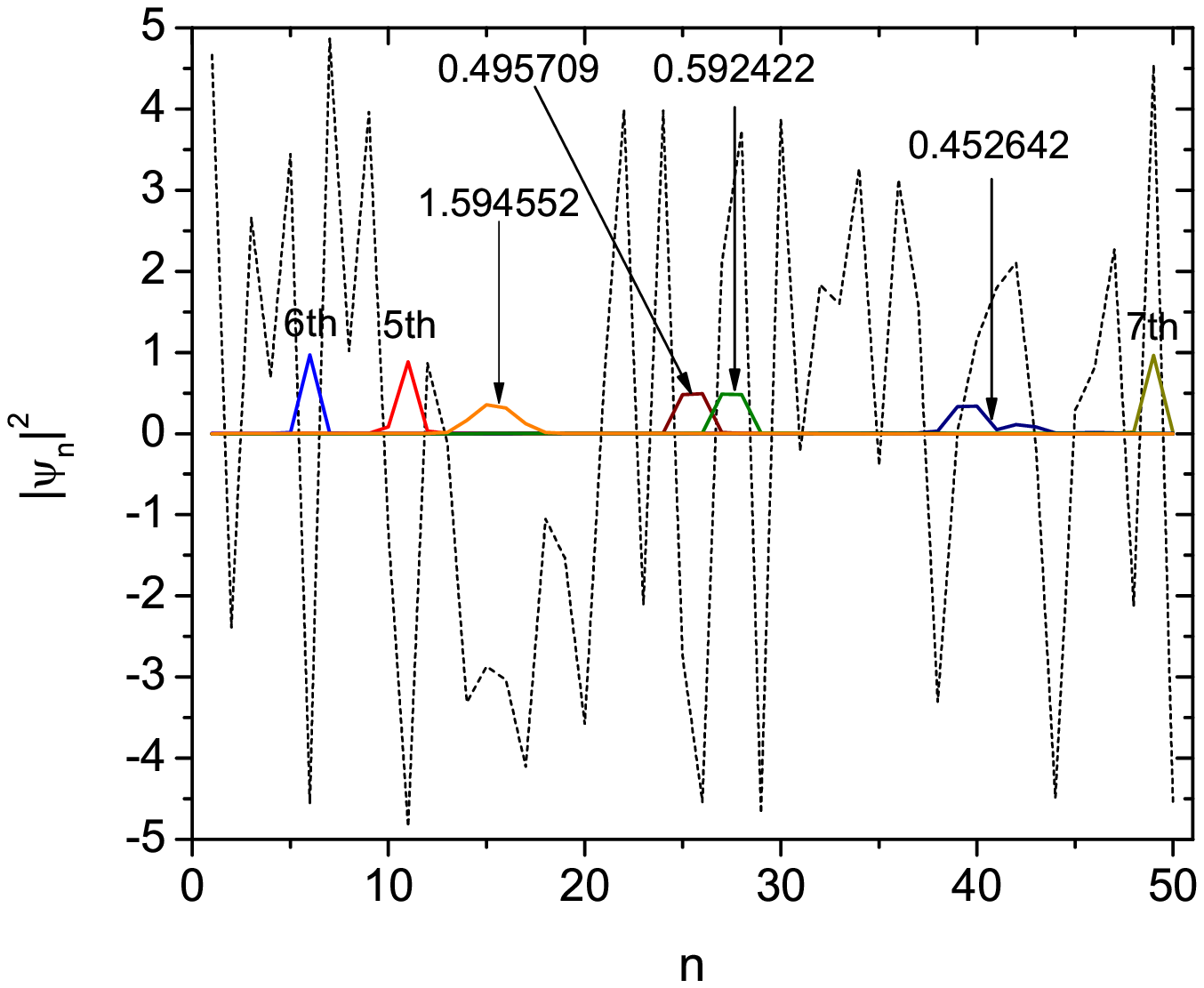}
	\caption{Spatial distributions of several typical wave functions for the eigenstates with zero and non-zero real parts of the energy are shown in the case of $N=50$ and $W=10$. The dashed curve indicates the imaginary part $\epsilon_{n}^{\rm I}$ of the on-site potential. It is clear that, as the disorder strength gets sufficiently large, the eigenstates with ${\rm Re}(E)= 0$ become more strongly localized than those with ${\rm Re}(E)\neq 0$.}
\end{figure}

It is worth mentioning that although it is not presented in this paper, we have checked numerically
that all the results reported above are also valid in the situations where the average value of the imaginary part of the on-site potential $\epsilon_{n}^{\rm I}$ takes any nonzero value. These include the cases where $\epsilon_{n}^{\rm I}$ takes only positive or negative random values.
The same results are obtained also in the cases where the real part of the on-site potential $\epsilon_{n}^{\rm R}$ takes an arbitrary constant (and nonrandom) value.

\section{Conclusion}
\label{sec:4}

In the present work, we have performed a numerical study of the localization properties in a 1D disordered lattice characterized by a non-Hermitian random Hamiltonian, where the on-site potential is random and imaginary. After obtaining the eigenstates and the corresponding eigenvalues by solving equation~(\ref{equation2}) numerically, we have calculated the averaged participation number, Shannon entropy and structural entropy as a function of other parameters. We have found that, in the presence of an imaginary random potential, all eigenstates are exponentially localized in the thermodynamic limit. Most importantly, we have found that anomalous Anderson localization occurs at the band center in the system under consideration. In contrast to the usual anomalies where a weaker localization is observed, our numerical results have clearly shown that the localization of the states at the band center is strongly enhanced. This phenomenon is associated with the occurrence of a large number of strongly-localized states with ${\rm Re}(E)=0$. We speculate that, in the strong-disorder regime, eigenstates with pure imaginary eigenvalues will become more strongly localized than all other eigenstates with non-zero real parts of the energy eigenvalue.
We hope that our results presented here will be an interesting addition to the study of disordered non-Hermitian systems.

\section*{Acknowledgments} This research is funded by Vietnam National Foundation for Science and Technology Development (NAFOSTED) under Grant No. 103.01-2018.05. It is also supported by a National Research Foundation of Korea Grant (NRF-2019R1F1A1059024) funded by the Korean Government.


\section*{References}
\begin{thebibliography}{10}

\bibitem {And}  Anderson P W 1958 {\it Phys. Rev.} {\bf 109} 1492

\bibitem {Lee} Lee P A and Ramakrishnan T V 1985 {\it Rev. Mod. Phys.} {\bf 57} 287

\bibitem {She} Sheng P (ed.) 1990 {\it Scattering and Localization of Classical Waves in Random Media} (Singapore: World Scientific)

\bibitem {Sch} Schwartz T, Bartal G, Fishman S and Segev M 2007 {\it Nature} {\bf 446} 52

\bibitem {Lah} Lahini Y, Avidan A, Pozzi F, Sorel M, Morandotti R, Christodoulides D N and Silberberg Y 2008 {\it Phys. Rev. Lett.} {\bf 100} 013906

\bibitem {Bil} Billy J, Josse V, Zuo Z, Bernard A, Hambrecht B, Lugan P, Cl\'{e}ment D, Sanchez-Palencia L, Bouyer P and Aspect A 2008 {\it Nature} {\bf 453} 891

\bibitem {Roa}  Roati G, D'Errico C, Fallani L, Fattori M, Fort C, Zaccanti M, Modugno G, Modugno M and Inguscio M 2008 {\it Nature} {\bf 453} 895

\bibitem {Aspect} Aspect A and Inguscio M 2009 {\it Phys. Today} {\bf 62} 30

\bibitem{Thouless1} Thouless D J 1979 {\it Percolation localization} in {\it Ill-Condensed Matter}, Balian R, Maynard R and Toulouse G (eds.) (Amsterdam: North-Hollan)

\bibitem{Czycholl} Czycholl G, Kramer B and MacKinnon A 1981 {\it Z. Phys. B} {\bf 43} 5

\bibitem{Kappus} Kappus M and Wegner F 1981 {\it Z. Phys. B} {\bf 45} 15

\bibitem{Kravtsov1} Kravtsov V E and Yudson V I 2010 {\it Phys. Rev. B} {\bf 82} 195120

\bibitem{Kravtsov2} Kravtsov V E and Yudson V I 2011 {\it Ann. Phys.} {\bf 326} 1672

\bibitem{Tessieri1} Tessieri L, Herrera-Gonz\'{a}lez I F and Izrailev F M 2012 {\it Physica E} {\bf 44} 1260

\bibitem{Nguyen1} Nguyen B P and Kim K 2012 {\it J. Phys.: Condens. Matter} {\bf 24} 135303

\bibitem{Ramola} Ramola K and Texier C 2014 {\it J. Stat. Phys.} {\bf 157} 497

\bibitem{Tessieri2}Tessieri L, Herrera-Gonz\'{a}lez I F and Izrailev F M 2015 {\it J. Phys. A: Math. Theor.} {\bf 48} 355001

\bibitem{Deych1}  Deych L I, Lisyansky A A and Altshuler B L 2000 {\it Phys. Rev. Lett.} {\bf 84} 2678

\bibitem{Deych2}  Deych L I, Lisyansky A A and Altshuler B L 2001 {\it Phys. Rev.} B {\bf 64} 224202

\bibitem{Hern}   Hern\'{a}ndez-Herrej\'{o}n J C, Izrailev F M and Tessieri L 2010 {\it J. Phys. A: Math. Theor.} {\bf 43} 425004

\bibitem{Derrida}  Derrida B and Gardner E 1984 {\it J. Phys. (Paris)} {\bf 45} 1283

\bibitem{Alloatti} Alloatti L 2009 {\it J. Phys.: Condens. Matter} {\bf 21} 045503

\bibitem{Sepehrinia1}  Sepehrinia R 2010 {\it Phys. Rev.} B {\bf 82} 045118

\bibitem{Sepehrinia2}  Sepehrinia R 2013 {\it J. Stat. Phys.} {\bf 153} 1039

\bibitem{Izrailev}  Izrailev F M, Krokhin A A and Makarov N M 2012 {\it Phys. Rep.} {\bf 512} 125

\bibitem{Krimer}  Krimer D O and Flach S 2010 {\it Phys. Rev. E} {\bf 82} 046221

\bibitem{Amir} Amir A, Hatano N and Nelson D R 2016 {\it Phys. Rev. E} {\bf 93} 042310

\bibitem{Fre} Freilikher V, Pustilnik M and Yurkevich I 1994 {\it Phys. Rev. B} {\bf 50} 6017

\bibitem{Paa}  Paasschens J C J, Misirpashaev T Sh and Beenakker C W J 1996 {\it Phys. Rev. B} {\bf 54} 11887

\bibitem{Asa} Asatryan A A, Nicorovici N A, Botten L C, Martijn de Sterke C, Robinson P A and McPhedran R C 1998 {\it Phys. Rev. B} {\bf 57} 13535

\bibitem{Jia} Jiang X, Li Q and Soukoulis C M 1999 {\it Phys. Rev. B} {\bf 59} R9007(R)

\bibitem {Hat} Hatano N and Nelson D R 1996 {\it Phys. Rev. Lett.} {\bf 77} 570

\bibitem {Sil} Silvestrov P G 2001 {\it Phys. Rev. B} {\bf 64} 075114

\bibitem{Kal} Kalish S, Lin Z and Kottos T 2012 {\it Phys. Rev. A} {\bf 85} 055802

\bibitem {Bas} Basiri A, Bromberg Y, Yamilov A, Cao H and Kottos T 2014  {\it Phys. Rev. A} {\bf 90} 043815

\bibitem{Vaz} V\'{a}zquez-Candanedo O, Hern\'{a}ndez-Herrej\'{o}n J C, Izrailev F M and Christodoulides D N 2014 {\it Phys. Rev. A} {\bf 89} 013832

\bibitem{Mej} Mej\'{i}a-Cort\'{e}s C and Molina M I 2015 {\it Phys. Rev. A} {\bf 91} 033815

\bibitem{Kar} Kartashov V, Hang C, Konotop V V, Vysloukh V A, Huang G and Torner L 2016 {\it Laser Photon. Rev.} {\bf 10} 100

\bibitem{Nguyen2} Nguyen B P and Kim K 2016 {\it Phys. Rev. A} {\bf 94} 062122

\bibitem{Christo} Christodoulides D N and Joseph R I 1988 {\it Opt. Lett.} {\bf 13} 794

\bibitem{Izyumov1} Izyumov A V and Simons B D 1999 {\it Europhys. Lett.} {\bf 45} 290

\bibitem{Izyumov2} Izyumov A V and Simons B D 1999 {\it Phys. Rev. Lett.} {\bf 83} 4373

\bibitem{Thouless} Thouless D J 1974 {\it Phys. Rep.} {\bf 13} 93

\bibitem{Wegner} Wegner F 1980 {\it Z. Phys. B} {\bf 36} 209

\bibitem{Izrailev2} Izrailev F M 1990 {\it Phys. Rep.} {\bf 196} 299

\bibitem{Oliveira} de Oliveira C R 2002 {\it Phys. Lett. A} {\bf 296} 165

\bibitem{Bogomolny} Bogomolny E and Giraud O 2011 {\it Phys. Rev. Lett.} {\bf 106} 044101

\bibitem{Casati2} Casati G, Guarneri I, Izrailev F, Fishman S and  Molinari L 1992 {\it J. Phys.: Condens. Matter} {\bf 4} 149

\bibitem{Pipek}  Pipek J and Varga I 1992 {\it Phys. Rev. A} {\bf 46} 3148

\end{thebibliography}
\end{document}